\documentclass[12pt]{article}
\usepackage{graphics}
\usepackage{bm}
\usepackage{graphicx}
\usepackage{amsmath}
\usepackage{amssymb}
\usepackage{amstext}
\usepackage{amscd}
\usepackage{amsfonts}
\usepackage{appendix}
\usepackage{color}
\newcommand{\nd}{\noindent}

\newcommand{\be}{\begin{equation}}
\newcommand{\ee}{\end{equation}}
\newcommand{\ben}{\begin{eqnarray}}
\newcommand{\een}{\end{eqnarray}}

\title{{\bf From the hypergeometric differential equation to a non linear  Schr\"{o}dinger one}}

\author{{A. Plastino$^1$, M.C.Rocca$^1$%  A.R. Plastino$^{2}$
} \\
\small{$^1$ La Plata National University and
Argentina's National Research Council}\\
\small{(IFLP-CCT-CONICET)-C. C. 727, 1900 La Plata - Argentina}\\
}
\date{\today}

\begin{document}

\maketitle

\begin{abstract}
We show that the q-exponential function  is a hypergeometric
function. Accordingly, it obeys the hypergeometric differential
equation. We demonstrate that this differential equation  can be
transformed into a non-linear Schr\"{o}dinger equation (NLSE).
This NLSE exhibits both similarities and differences vis-a-vis the
Nobre-Rego Monteiro-Tsallis one. \vskip 3mm

\nd {\bf Keywords:} Non linear Schr\"{o}dinger equations,
separation of variables, hypergeometric function.

\end{abstract}

\newpage

\renewcommand{\theequation}{\arabic{section}.\arabic{equation}}

\section{Introduction}

\nd In 2011,  Nobre, Rego-Monteiro and Tsallis (NRT)
\cite{NRT11,NRT12, NRT13,NRT14,NRT15} introduced an intriguing  new version of the
nonlinear Schr\"odinger equation (NLSE), an interesting proposal
that one may regard as  part of a project to explore non-linear
versions of some of the fundamental equations of physics, a
research venue actively visited in  recent times \cite{S07,SS99}.
Earlier
 non-linear versions of the SE  have found
 application in diverse areas (fiber optics and water waves, for instance) \cite{SS99}.  A  most
 studied NLSE involves a cubic nonlinearity in the wave function. In
quantum settings the NLSE usually rules the behavior of a
single-particle's wave function that, in turn,  provides an
effective, mean-field description of a quantum many-body system.
 An important case is the Gross-Pitaevskii equation, employed in researching
 Bose-Einstein condensates \cite{PS03}. The cubic nonlinear term
appearing in the  Gross-Pitaevskii equation describes short-range
interactions between   the condensate's constituents. The NLSE for
the system's (effective) single-particle wave function is found
assuming a Hartree-Fock-like form for the global many-body wave
function, with a Dirac's delta form for the inter-particle
potential.

\nd The NRT equation derives from the thermo-statistical formalism
based upon the Tsallis $S_q$ non-additive, power-law information
measure.  Applications of the functional  $S_q$  involve diverse
physical systems and processes,  having attracted much attention
in the last 20 years (see, for example,
\cite{GT04,T09,N11,AdSMNC10,B09, O1,O2,O3,O4}, and references
therein). In particular, the $S_q$ entropy has proved to be useful
for the analysis of diverse problems in quantum physics
\cite{MPP2012,ZP10,SAA10,VPPD12,MML02,ZP06,CAFA11,SSAA12,TBD98}.

\nd In this paper we traverse a totally different road. We start
from the differential equation that governs hypergeometric
functions and derive from it a new NLSE that is different from,
but exhibits some similarities with, the NRT.

\setcounter{equation}{0}

\section{A New Non-Linear Schr\"{o}dinger Equation}

\setcounter{equation}{0}

\nd The   q-exponential $e_q$ is  defined as
$e_q(x)=[1+(q-1)x]_+^{\frac {1} {1-q}}$, that is,

\ben & e_q(x)=[1+(q-1)x]_+^{\frac {1} {1-q}}= [1+(q-1)x]^{\frac
{1} {1-q}}\;\;{\rm if}\;\; 1+(q-1)x>0 \cr & e_q(x)=0,
\;\;\;\;\;{\rm otherwise} \;\;\;\;\;({\rm with} \,\,q\in
\mathcal{R}).\een A search in  \cite{tp1} reveals that
\begin{equation}
\label{eq2.1} F(-\alpha,\gamma;\gamma;-z)=(1+z)^{\alpha},
\end{equation}
which yields for $e_q[(i/\hbar)(px-Et) ] $ the relation (with
$E=\frac {p^2} {2m}$)
\begin{equation}
\label{eq2.2} \left[1+\frac {i} {\hbar}(1-q)(px-Et)\right]^{\frac
{1} {1-q}}= F\left[\frac {1} {q-1},\gamma;\gamma;\frac {i} {\hbar}
(q-1)(px-Et)\right],
\end{equation}  which is a fundamental result for us.
Now, according to \cite{tp2}, the hypergeometric function obeys
the following,  differential equation (primes denote derivatives
with respect to $z$) \ben \label{eq2.3} &
z(1-z)F''(\alpha,\beta;\gamma;z)+\cr & +
[\gamma-(\alpha+\beta+1)z]F'(\alpha,\beta;\gamma;z)- \alpha\beta
F(\alpha,\beta;\gamma;z)=0, \een so that, specializing things for
our instance (\ref{eq2.2}) we encounter
\[\frac {i} {\hbar}(q-1)(px-Et)\left[
1-\frac {i} {\hbar}(q-1)(px-Et)\right]
F''\left[\frac {1} {q-1},\gamma;\gamma;\frac {i} {\hbar}
(q-1)(px-Et)\right]+\]
\[\left[\gamma-\left(\frac {1} {q-1}+\gamma+1\right)
\frac {i} {\hbar}(q-1)(px-Et)\right]
F'\left[\frac {1} {q-1},\gamma;\gamma;\frac {i} {\hbar}
(q-1)(px-Et)\right]-\]
\begin{equation}
\label{eq2.4} \frac {\gamma} {q-1} F\left[\frac {1}
{q-1},\gamma;\gamma;\frac {i} {\hbar} (q-1)(px-Et)\right]=0.
\end{equation}
 This allows for a relation between the  derivative with respect to the
 argument and the partial derivative with respect to time, for this   hypergeometric function
\[F'\left[\frac {1} {q-1},\gamma;\gamma;\frac {i} {\hbar}
(q-1)(px-Et)\right]=\]
\begin{equation}
\label{eq2.5} \frac {i\hbar} {(q-1)E}\frac {\partial} {\partial t}
F\left[\frac {1} {q-1},\gamma;\gamma;\frac {i} {\hbar}
(q-1)(px-Et)\right].
\end{equation}
In analogous fashion we obtain, for the second partial derivative
with respect to the position
\[F''\left[\frac {1} {q-1},\gamma;\gamma;\frac {i} {\hbar}
(q-1)(px-Et)\right]=\]
\begin{equation}
\label{eq2.6} -\frac {\hbar^2} {(q-1)^2p^2}\frac {\partial^2}
{\partial x^2} F\left[\frac {1} {q-1},\gamma;\gamma;\frac {i}
{\hbar} (q-1)(px-Et)\right].
\end{equation}
Replacing  (\ref{eq2.5}) and(\ref{eq2.6}) into (\ref{eq2.4}), this
last equation adopts the appearance
\[-\frac {i} {\hbar}(q-1)(px-Et)\left[
1-\frac {i} {\hbar}(q-1)(px-Et)\right]\times\]
\[\frac {\hbar^2} {(q-1)^2p^2}\frac {\partial^2} {\partial x^2}
F\left[\frac {1} {q-1},\gamma;\gamma;\frac {i} {\hbar}
(q-1)(px-Et)\right]+\]
\[\left[\gamma-\left(\frac {1} {q-1}+\gamma+1\right)
\frac {i} {\hbar}(q-1)(px-Et)\right]\times\]
\[\frac {i\hbar} {(q-1)E}\frac {\partial} {\partial t}
F\left[\frac {1} {q-1},\gamma;\gamma;\frac {i} {\hbar}
(q-1)(px-Et)\right]\]
\begin{equation}
\label{eq2.7} -\frac {\gamma} {q-1} F\left[\frac {1}
{q-1},\gamma;\gamma;\frac {i} {\hbar} (q-1)(px-Et)\right]=0,
\end{equation}
that can be recast in the fashion
\[-\frac {i} {\hbar}(q-1)(px-Et)\left[
1-\frac {i} {\hbar}(q-1)(px-Et)\right]\times\]
\[\frac {\hbar^2} {(q-1)^2m^2}\frac {\partial^2} {\partial x^2}
F\left[\frac {1} {q-1},\gamma;\gamma;\frac {i} {\hbar}
(q-1)(px-Et)\right]+\]
\[\left[\gamma-\left(\frac {1} {q-1}+\gamma+1\right)
\frac {i} {\hbar}(q-1)(px-Et)\right]\times\]
\[i\hbar\frac {\partial} {\partial t}
F\left[\frac {1} {q-1},\gamma;\gamma;\frac {i} {\hbar}
(q-1)(px-Et)\right]\]
\begin{equation}
\label{eq2.8} -\gamma(q-1)E F\left[\frac {1}
{q-1},\gamma;\gamma;\frac {i} {\hbar} (q-1)(px-Et)\right]=0.
\end{equation}
Deriving (\ref{eq2.2})  with respect to time we obtain:
\[-\gamma E
F\left[\frac {1} {q-1},\gamma;\gamma;\frac {i} {\hbar}
(q-1)(px-Et)\right]=\]
\[-\frac {i\hbar} {q-1}\gamma
\left\{F\left[\frac {1} {q-1},\gamma;\gamma;\frac {i} {\hbar}
(q-1)(px-Et)\right]\right\}^{(1-q)}\times\]
\begin{equation}
\label{eq2.9} \frac {\partial} {\partial t} F\left[\frac {1}
{q-1},\gamma;\gamma;\frac {i} {\hbar} (q-1)(px-Et)\right].
\end{equation}
For simplicity, let us abbreviate
\begin{equation}
\label{eq2.10} F\equiv F\left[\frac {1} {q-1},\gamma;\gamma;\frac
{i} {\hbar} (q-1)(px-Et)\right].
\end{equation}
Using now  (\ref{eq2.9}), Eq.  (\ref{eq2.8}) becomes
\[-\frac {\hbar^2} {2m(q-1)}
\left[1-F^{(1-q)}\right]F^{(1-q)}
\frac {\partial^2} {\partial x^2}F+\]
\[i\hbar\left\{\gamma+\left(\frac {1} {q-1}+
\gamma+1\right)\left[F^{(1-q)}-1\right]\right\}
\frac {\partial} {\partial t} F-\]
\begin{equation}
\label{eq2.11} i\hbar\gamma F^{(1-q)}\frac {\partial} {\partial t}
F=0.
\end{equation}
Simplifying things in this last relation we arrive at
\begin{equation}
\label{eq2.12} -\frac {\hbar^2} {2m} F^{(1-q)} \frac {\partial^2}
{\partial x^2}F-i\hbar q\frac {\partial} {\partial t}F=0,
\end{equation}
that can be rewritten as
\begin{equation}
\label{eq2.13} i\hbar q\frac {\partial}{\partial t}F= F^{(1-q)}H_0
F,
\end{equation}
where $H_0$ is the free particle Hamiltonian, Note that, for
$q=1$, one reobtains  Schr\"{o}dinger's free particle equation.
Now, if instead of  (\ref{eq2.2}) we deal just with
\begin{equation}
\label{eq2.14} F(x,t)=A \left[1+\frac {1}
{\hbar}(1-q)(px-Et)\right]^{\frac {1} {1-q}},
\end{equation}
then $F(0,0)=A$ and  (\ref{eq2.13}) becomes
\begin{equation}
\label{eq2.15} i\hbar q\frac {\partial}{\partial t} \left[\frac
{F(x,t)} {F(0,0)}\right]= {\left[\frac {F(x,t)}
{F(0,0)}\right]}^{(1-q)}H_0 \left[\frac {F(x,t)} {F(0,0)}\right],
\end{equation}
or, equivalently,
\begin{equation}
\label{eq2.16} i\hbar q {\left[\frac {F(x,t)}
{F(0,0)}\right]}^{(q-1)} \frac {\partial}{\partial t} \left[\frac
{F(x,t)} {F(0,0)}\right]= H_0 \left[\frac {F(x,t)}
{F(0,0)}\right],
\end{equation}
that, in turn can be recast as
\begin{equation}
\label{eq2.17} i\hbar \frac {\partial}{\partial t} \left[\frac
{F(x,t)} {F(0,0)}\right]^q= H_0 \left[\frac {F(x,t)}
{F(0,0)}\right].
\end{equation}
At this stage we realize that this last equation could be
'generalized' to any Hamiltonian  $H$ as
\begin{equation}
\label{eq2.18} i\hbar \frac {\partial}{\partial t} \left[\frac
{\psi(x,t)} {\psi(0,0)}\right]^q= H \left[\frac {\psi(x,t)}
{\psi(0,0)}\right].
\end{equation}
With the change of variables  $[\psi(x,t)]^q=\phi(x,t)$, Eq.
(\ref{eq2.18}) takes the form
\begin{equation}
\label{eq2.19} i\hbar\frac {\partial}{\partial t} \left[\frac
{\phi(x,t)} {\phi(0,0)}\right]=H {\left[\frac {\phi(x,t)}
{\phi(0,0)}\right]}^{\frac {1} {q}},
\end{equation}
which trivially reduces to the ordinary Schr\"odinger equation for
$q=1$.

\section{Separation of Variables and free particle case}

\setcounter{equation}{0}

Consider now Eq.  (\ref{eq2.19}) for a time-independent $H$. A
separable situation ensues. Let
\begin{equation}
\label{eq3.1}
\frac {\psi (x,t)}{\psi(0,0)}=\frac {f(t)}{f(0)}\frac {g({x})}{g(0)}.\end{equation}
Then, (\ref{eq2.19}) becomes
\begin{equation}
\label{eq3.2}
i\hbar
\left[\frac {g({x})} {g(0)}\right]^q
\frac {d} {d t}
\left[\frac {f(t)} {f(0)}\right]^q=
\left[\frac {f(t)} {f(0)}\right]
H\left[\frac {g({x})} {g(0)}\right].
\end{equation}
Rewrite  (\ref{eq3.2})  as

\begin{equation}
\label{eq3.3}
i\hbar
\left[\frac {f(t)} {f(0)}\right]^{-1}
\frac {d} {d t}
\left[\frac {f(t)} {f(0)}\right]^q=
\left[\frac {g({x})} {g(0)}\right]^{-q}
H\left[\frac {g({x})} {g(0)}\right]=\lambda,
\end{equation}
from which we gather that
 $\lambda= constant$ and

\begin{equation}
\label{eq3.4}
i\hbar
\frac {d} {d t}
\left[\frac {f(t)} {f(0)}\right]^q=\lambda
\left[\frac {f(t)} {f(0)}\right],
\end{equation}
\begin{equation}
\label{eq3.5}
H\left[\frac {g({x})} {g(0)}\right]=\lambda
\left[\frac {g(x)} {g(0)}\right]^q.
\end{equation}
Indeed, spatial and temporal variables have been decoupled.
Pass now to the free-particle case. Set in
 (\ref{eq3.4}) - (\ref{eq3.5})  $\psi(0,0)=1$. We have

\begin{equation}
\label{eq3.6}
i\hbar
\frac {d} {dt}
\left[f(t)\right]^q=\lambda f(t)
\end{equation}
\begin{equation}
\label{eq3.7}
-\frac {\hbar^2} {2m}
\frac {d} {dx^2}
g(x)=\lambda
\left[g(x)\right]^q
\end{equation}
It is straightforward to ascertain that a possible solution is
\begin{equation}
\label{eq3.8}
E=\frac {p^2} {2m}=\lambda
\end{equation}
\begin{equation}
\label{eq3.9}
f(t)=\left[1+\frac {i} {\hbar}\frac {(1-q)} {q}
Et\right]^{\frac {1} {q-1}}
\end{equation}
\begin{equation}
\label{eq3.10}
g(x)=\left[1+\frac {i} {\hbar}\frac {(1-q)} {\sqrt{2(q+1)}}
px\right]^{\frac {2} {1-q}}
\end{equation}
 We have   two  free-particle solutions.
For $q\rightarrow 1$ both solutions tend to the usual solution
for the habitual Schr\"{o}dinger  equation.

\section{The NRT Equation}

\setcounter{equation}{0}

For comparison purposes, we remember that  NRT
equation reads \cite{NRT11,NRT12}
\begin{equation}
\label{eq4.1}
i\hbar (2-q)
\frac {\partial}{\partial t}
\left[\frac {\psi(\vec{x},t)} {\psi(0,0)}\right]=
H_0
\left[\frac {\psi(\vec{x},t)} {\psi(0,0)}\right]^{2-q}.
\end{equation}
{\it We} introduce here the change of variables  $\phi=\psi^{2-q}$ to obtain

\begin{equation}
\label{eq4.2} i\hbar (2-q) \frac {\partial}{\partial t}
\left[\frac {\phi(\vec{x},t)} {\phi(0,0)}\right]^{ \frac {1}
{2-q}}= H_0 \left[\frac {\phi(\vec{x},t)} {\phi(0,0)}\right].
\end{equation}
We thus see that our present equation is, for the free particle
 case, equivalent to the NRT equation.

\nd The last  NRT equation is amenable of generalization. Thus,
for an arbitrary Hamiltonian $H$ one may write
\begin{equation}
\label{eq4.3} i\hbar (2-q) \frac {\partial}{\partial t}
\left[\frac {\phi({x},t)} {\phi(0,0)}\right]^{ \frac {1} {2-q}}= H
\left[\frac {\phi({x},t)} {\phi(0,0)}\right],
\end{equation}
so that \be \label{eq4.4} i\hbar (2-q) \frac {\partial}{\partial
t} \left[\frac {\psi (x,t)} {\psi(0,0)}\right]= H \left[\frac
{\psi(x,t)} {\psi(0,0)}\right]^{2-q}.\ee We show next that
NRT-separation of variables is feasible.

\section{NRT separation of variables}

\setcounter{equation}{0}

We show that, for a time independent $H$, (\ref{eq4.4}) can be
separated. For this set
\begin{equation}
\label{eq5.1} \frac {\psi({x},t)} {\psi(0,0)}= \frac {f(t)} {f(0)}
\frac {g({x})} {g(0)},
\end{equation}
so that  (\ref{eq4.4}) becomes
\begin{equation}
\label{eq5.2} i\hbar (2-q) \left[\frac {g({x})} {g(0)}\right]
\frac {d} {d t} \left[\frac {f(t)} {f(0)}\right]= \left[\frac
{f(t)} {f(0)}\right]^{2-q} H\left[\frac {g({x})}
{g(0)}\right]^{2-q},
\end{equation}
that can be rewritten as

\begin{equation}
\label{eq5.3} i\hbar \left[\frac {f(t)} {f(0)}\right]^{q-2} \frac
{d} {d t} \left[\frac {f(t)} {f(0)}\right]= \left[\frac {g({x})}
{g(0)}\right]^{-1} H\left[\frac {g({x})}
{g(0)}\right]^{2-q}=\lambda,
\end{equation}
with  $\lambda= constant$. Accordingly:

\begin{equation}
\label{eq5.4} i\hbar (2-q) \frac {d} {d t} \left[\frac {f(t)}
{f(0)}\right]=\lambda \left[\frac {f(t)} {f(0)}\right]^{2-q},
\end{equation}
\begin{equation}
\label{eq5.5} H\left[\frac {g({x})} {g(0)}\right]^{2-q}=\lambda
\left[\frac {g({x})} {g(0)}\right].
\end{equation}
In the free particle case one has
\begin{equation}
\label{eq5.6} i\hbar (2-q) \frac {d} {dt} f(t)=\lambda
[f(t)]^{2-q},
\end{equation}
\begin{equation}
\label{eq5.7} -\frac {\hbar^2} {2m} \frac {d} {dx^2}
[g(x)]^{2-q}=\lambda g(x),
\end{equation}
with solution

\begin{equation}
\label{eq5.8} E=\frac {p^2} {2m}=\lambda,
\end{equation}
\begin{equation}
\label{eq5.9} f(t)=\left[1+\frac {i} {\hbar}\frac {(1-q)} {2-q}
Et\right]^{\frac {1} {q-1}},
\end{equation}
\begin{equation}
\label{eq5.10} g(x)=\left[1+\frac {i} {\hbar}\frac {(1-q)}
{\sqrt{2(2-q)(3-q)}} px\right]^{\frac {2} {1-q}}.
\end{equation}
This solution does NOT coincide  with  the solution obtained in
Section 3, except for the case $q\rightarrow 1$.

\section{Conclusions}

\nd We have noticed, first of all,  an important feature of the
q-exponential function, namely,

\begin{equation}
\label{eq2.Concl} \left[1+\frac {i}
{\hbar}(1-q)(px-Et)\right]^{\frac {1} {1-q}}= F\left[\frac {1}
{q-1},\gamma;\gamma;\frac {i} {\hbar} (q-1)(px-Et)\right],
\end{equation}
that it is an hypergeometric function. From such result, it is
clear that the q-exponential function obeys the  hypergeometric
differential equation (2.4).

\nd Now, suitably manipulating equation (2.4) we have reached a
non linear Schr\"odinger equation that resembles the NRT one
introduced in \cite{NRT11}. We have seen that, for time
independent Hamiltonians, separation of spatial from temporal
variables ensues, as in the ordinary Schr\"odinger equation. We
have also seen that this separation can be made in the NRT case.

\nd It is worth mentioning that, in  contexts unrelated to the
present one, connections between hypergeometric functions and
q-statistics' applications have been encountered. See, for
examples, \cite{t1,t2}.

 \setcounter{equation}{0}

\section{Acknowledgements}

\setcounter{equation}{0}

\nd We gratefully acknowledge Prof.  A. R. Plastino
for useful suggestions and comments.

\end{document}